# Agents Play Mix-game

Chengling Gou

Physics Department, Beijing University of Aeronautics and Astronautics
37 Xueyuan Road, Haidian District, Beijing, China, 100083

Physics Department, University of Oxford
Clarendon Laboratory, Parks Road, Oxford, OX1 3PU, UK
gouchengling@hotmail.com, c.gou1@physics.ox.ac.uk

Abstract: In mix-game which is an extension of minority game, there are two groups of agents; group1 plays the majority game, but the group2 plays the minority game. This paper studies the change of the average winnings of agents and volatilities vs. the change of mixture of agents in mix-game model. It finds that the correlations between the average winnings of agents and the mean of local volatilities are different with different combinations of agent memory length when the proportion of agents in group 1 increases. This study result suggests that memory length of agents in group1 be smaller than that of agent in group2 when mix-game model is used to simulate the financial markets.

1. **Introduction:**

Challet and Zhang's MG model, together with the original bar model of Arthur [1], attracts a lot of following studies. Given the MG's richness and yet underlying simplicity, the MG has also received much attention as a financial market model [2]. The MG comprises an odd number of agents choosing repeatedly between the options of buying (1) and selling (0) a quantity of a risky asset. The agents continually try to make the minority decision i.e. buy assets when the majority of other agents are selling and sell when the majority of other agents are buying. Neil F. Johnson [3, 4] and coworkers extended MG by allowing a variable number of active traders at each timestep--- they called their modified game as the Grand Canonical Minority Game (GCMG). The GCMG, and to a lesser extent the basic MG itself, can reproduce the stylized facts of financial markets, such as volatility clustering and fat-tail distributions. However, there are some weaknesses in MG and GCMG. First, the diversity of agents is limited, since agents all have the same memory and time-horizon. Second, in real markets, some agents are tendency-followers, i.e. "noise traders" [5, 6, 7, 8, 9, 10, 11, 12], who effectively play a majority game; while others are "foundation traders", who effectively play a minority game.

In order to create an agent-based model which more closely mimics a real financial market, I proposed a mix-game model which is a modification of MG [13]. In mix-game model there are two groups of agents: each group has different memory and time-horizon. The most important modification is to make one group plays the minority game and the other plays the majority game. Through simulations, I find out that the fluctuations of local volatilities change a lot by adding some agents who play majority game into MG, but the stylized features of MG don't change obviously except agents with memory length 1 and 2. I also give suggestions about how to use mix-game to model financial markets and show the example of modeling Shanghai stock market by means of mix-game model [13].

In this paper, I further examine the correlations between the average winnings of agents and the local volatilities of systems in mix-game model when the proportion of agents in group1 increases from 0 to 0.4. In section 2, I describe the mix-game model and the simulation conditions. In section 3, the simulation results and discussion are presented. In section 4, I calculate the quantitative correlations. In section 5, the conclusion is reached.

**2. Simulation:**

Mix-game model is an extension of minority game (MG), so its structure is similar to MG.

In mix-game, there are two groups of agents; group1 plays the majority game, and the group2 plays the minority game. N (odd number) is the total number of the agents and N1 is number of agents in group1. The system resource is r = N*L, where L<1 is the proportion of resource of the system. All agents compete in the system for the limited resource r. T1 and T2 are the time horizon lengths of the two groups, and m1 and m2 denote the memory lengths of the two groups, respectively.

The global information only available to the agents is a common bit-string "memory" of the m1 or m2 most recent competition outcomes (1 or 0). A strategy consists of a response, i.e., 0 (sell) or 1 (buy), to each possible bit string; hence there are $2^{2^{m1}}$ or $2^{2^{m2}}$ possible strategies for group1 or group2, respectively, which form full strategy spaces (FSS). At the beginning of the game, each agent is assigned s strategies and keeps them unchangeable during the game. After each turn, agents assign one (virtual) point to a strategy which would have predicted the correct outcome. For agents in group1, they will reward their strategies one point if they are in the majority; for agents in group2, they will reward their strategies one point if they are in the minority. Agents collect the virtual points for their strategies over the time horizon T1 or T2, and they use their strategies which has the highest virtual point in each turn. If there are two strategies which have the highest virtual point, agents use coin toss to decide which strategy to be used. Excess demand is equal to the number of ones (buy) which agents choose minus the number of zeros (sell) which agents choose. According to a widely accepted assumption that excess demand exerts a force on the price of the asset and the change of price is proportion to the excess demand in a financial market [14, 15, 16], the time series of price of the asset can be calculated based on the time series of excess demand.

In simulation, the distribution of initial strategies of agents is randomly distributed and keeps unchanged during the games. Simulation turns are 3000. The window length of local volatility is 5. Total number of agents is 201. Number of strategies per agent is 2.

## 3. Simulation results and discussions

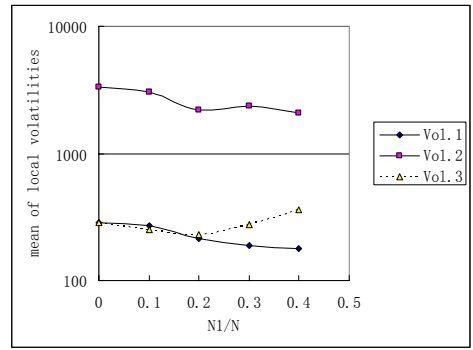

Fig.1 means of local volatilities vs. different N1/N, Vol.1 representing mean of local volatilities of m1=m2=6, T1=T2=60; Vol.2 representing mean of local volatilities of m1=6, m2=3, T1=60, T2=12; Vol.3 representing mean of local volatilities of m1=3, m2=6, T1=12, T2=60.

Fig.1 shows that means of local volatilities (Vol.1 and Vol.2) decrease while N1/N increases from 0 to 0.4 under condition of m1=m2=6, T1=T2=60 and condition of m1=6, m2=3, T1=60, T2=12. But under condition of m1=3, m2=6, T1=12, T2=60, mean of local volatilities (Vol.3) has a minimum value at N1/N=0.2.

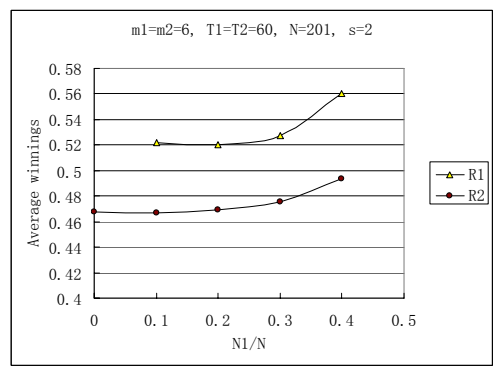

Fig.2 average winnings per agent per turn for mix-game vs. different proportion of agents in group1 when m1=m2=6, T1=T2=60; R1 represents the average winning per agent per turn of group1, and R2 represents the average winning per agent per turn of group2.

From Fig. 2, one can find that the average winnings (R1 and R2) of these two groups increase when N1/N is larger than 0.1 and the average winning of group 1 is larger than that of group 2. Agents in both groups benefit from the increase of the number of agents in group1. Comparing Fig.2 with Fig.1, one can find that the mean of local volatilities (Vol.1) decreases accompanying with the increase of the average winnings of group1 and group2 (R1, R2) while N1/N increases from 0 to 0.4 under the condition of m1=m2=6, T1=T2=60. This means that the improvement of the efficiency of systems is accordant with the improvement of the performance of individual agents under this simulation condition. This result shows that both the efficiency of systems and the performance of individual agents increase while the proportion of agents in group 1 increases. Similar phenomena can be found in ecological systems, computing systems and economic systems in which agents (species, computing tasks and firms) having different niches will improve the efficiencies of systems and their own performance [17].

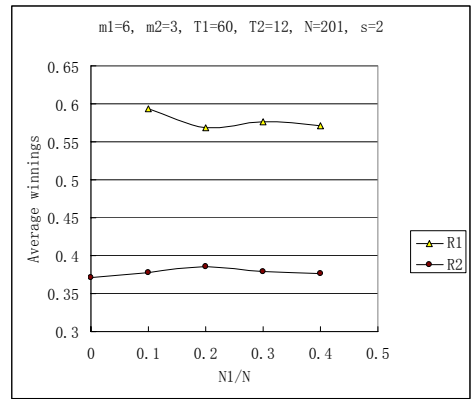

Fig.3 average winnings per agent per turn for mix-game vs. different proportion of agents in group1 when m1=6, m2=3, T1=60, T2=12. R1, R2 have the same meaning as that indicated in Fig.2.

From Fig. 3, one can find that the average winnings (R1 and R2) of these two groups don't change obviously when N1/N increases from 0 to 0.4, except R1 at N1/N=0.1. The average winnings of group 1 are larger than those of group 2. Comparing Fig.3 with Fig.1, one can find that the mean of local volatilities (Vol.2) decreases while N1/N increases from 0 to 0.4 under the condition of m1=6, m2=3, T1=60, T2=12, but the change of the average winnings of group1 and group2 (R1, R2) seems more complicated than that in Fig.2.

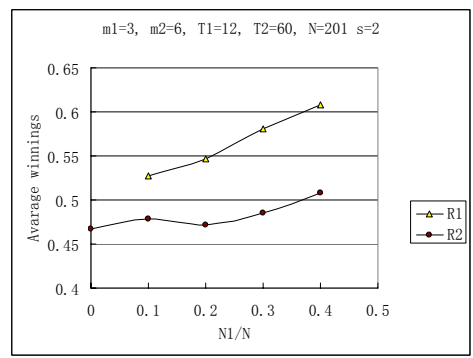

Fig.4 average winnings per agent per turn for mix-game vs. different proportion of agents in group1 when m1=3, m2=6, T1=12, T2=60. R1, R2 have the same meaning as that indicated in Fig.2.

From Fig. 4, one can find that the average winnings (R1 and R2) of these two groups increase obviously when N1/N increases from 0 to 0.4 and the average winning of group 1 (R1) is larger than that of group 2 (R2). Agents in both groups benefit from the increase of the number of agents in group1. Comparing Fig.4 with Fig.1, one can find that the mean of local volatilities (Vol.3) decreases slightly when N1/N increase from 0 to 0.2, then it increase while N1/N increases from 0.2 to 0.4 under the condition of m1=3, m2=6, T1=12, T2=60, accompanying with the increase of the average winnings of group1 and group2 (R1, R2). This means that the improvement of the performance of individual agents accompanies with the decrease of the system efficiency under this simulation condition. Agents can make profits from the larger fluctuation of systems, which is accordant with the reality of financial markets.

For example, the combinations of parameters in this simulation include the configuration of parameters which is used to model the Shanghai Index [13]. For Shanghai Index, there are two suitable configurations of parameters: m1=3, T1=12, m2=6, T2=60, N1=40, N=201 and m1=4, m2=6, T1=T2=12, N=201, N1=72, respectively. Fig.5 shows the log-log plot of Shanghai-daily absolute returns and Fig.6 shows the log-log plot of mix-game absolute returns of parameters of m1=3, T1=12, m2=6, T2=60, N1=40, N=201, s=2. These two figures look very similar. This implies that we need to make m1 smaller than m2 when we mimic financial markets by means of mix-game model.

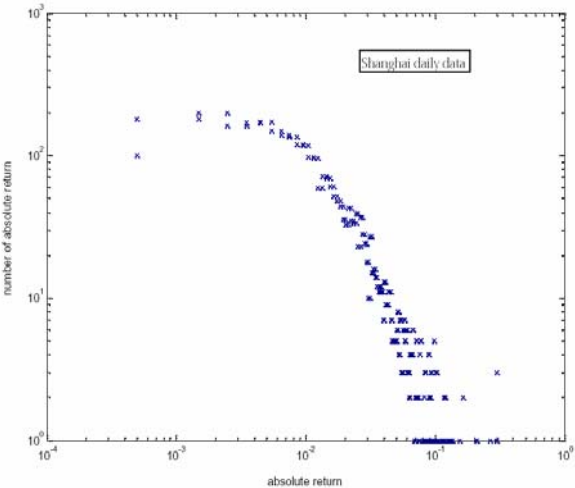 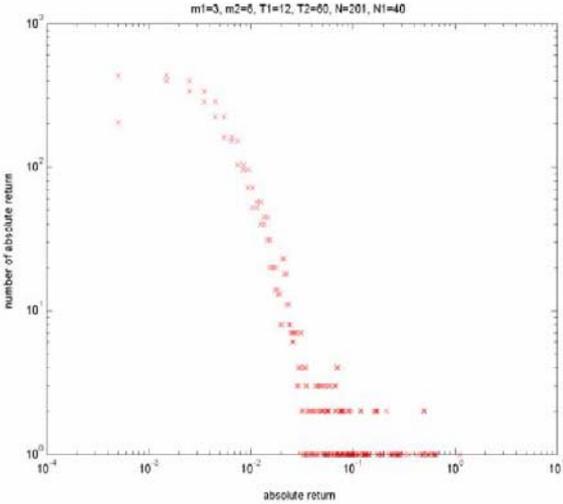

Fig.5 log-log plot of Shanghai-daily absolute returns    Fig.6 log-log plot of mix-game absolute returns

## 4. Calculation of correlations

Table 1 correlations of R1, R1 and Vol.1 under the condition of m1=m2=6, T1=T2=60

| Correlation | *R1* | *R2* | *Vol. 1* |
|---|---|---|---|
| R1 | 1 | | |
| R2 | 0.98 | 1 | |
| Vol. 1 | −0.63 | −0.76 | 1 |

Table 2 correlations of R1, R1 and Vol.2 under the condition of m1=6, m2=3, T1=60, T2=12

| Correlation | *R1* | *R2* | *Vol. 2* |
|---|---|---|---|
| R1 | *1* | | |
| R2 | *−0.48* | *1* | |
| Vol. 2 | *0.98* | *−0.67* | 1 |

Table 3 correlations of R1, R1 and Vol.3 under the condition of m1=3, m2=6, T1=12, T2=60,

| Correlation | *R1* | *R2* | *Vol. 3* |
|---|---|---|---|
| R1 | 1 | | |
| R2 | 0.87 | 1 | |
| Vol. 3 | 0.89 | 0.82 | 1 |

Table 1, 2 and 3 give the quantitative results of correlations among R1, R2 and the means of local volatilities under these three configurations of parameters. The results in table 1 and table 3 are accordant with the qualitative analysis in section 3 and show that the correlations between R1 and R2 in both situations are positive. Table 2 gives a clearer picture about the correlations among R1, R2 and Vol.3 under the condition of m1=3, m2=6, T1=12, T2=60; the correlation between R1 and R2 is negative; the correlation between R1 and Vol.3 is positive while the correlation between R2 and Vol.3 is negative.

## 5. Conclusion

The correlations between the average winnings of agents (R1 and R2) and the mean of local volatilities are different with different combinations of m1 and m2 when the proportion of agents in group 1 increases: the correlations are negative with parameters of m1=m2=6,

T1=T2=60, N=201 and s=2; under the condition of m1=6, m2=3, T1=60, T2=12, N=201 and s=2, the correlation of variable pair (R1 and Vol.2) is negative while the correlation of variable pair (R2 and Vol.2) is positive; the correlations are positive with parameters of m1=3, m2=6, T1=12, T2=60, N=201 and s=2.

This study result suggests that m1 be smaller than m2 when mix-game model is used to simulate the financial markets.

**Acknowledgements**

This research is supported by Chinese Overseas Study Fund. Thanks Professor Neil F. Johnson for suggesting modification of agents' memories and discussing about the calculation of average winning of agents. Thanks David Smith for providing the original program code of MG.

**Reference**
1. W.B. Arthur, *Science* 284, 107 (1999).
2. D. Challet, and Y. C. Zhang, *Phyisca A* 246, 407(1997);
3. Neil F. Johnson, Paul Jefferies, and Pak Ming Hui, *Financial Market Complexity*, Oxford Press(2003);
4. Paul Jefferies and Neil F. Johnson, *Designing agent-based market models,* Oxford Center for Computational Finance working paper: OCCF/010702;
5. T. Lux, *Herd Behavior, Bubbles and Crashes.* Economic Journal 105(431): 881-896(1995).
6. T. Lux, and M. Marchesi *Scaling and criticality in a stochastic multi-agent model of a financial market.* Nature 397(6719): 498-500 (1999)."
7. J. V. Andersen, and D. Sornette, *The $-game*, cond-mat/0205423;
8. Challet, Damien, *Inter-pattern speculation: beyond minority, majority and $-games,* arXiv: physics/0502140 v1.
9. F. Slanina and Y.-C. Zhang, Physica A 289, 290 (2001).
10. Yoon, Seong-Min, and Kim, Kyungsik, *Dynamical Minority Games in Futures Exchange Markets*, arXiv: physics/0503016 v1.
11. J.V. Andersen and D. Sornette, Eur. Phys. J. B 31, 141 (2003).
12. I. Giardina and J.-P. Bouchaud, Eur. Phys. J. B 31, 421 (2003).
13. Chengling Gou, *Dynamic Behaviors of Mix-game Models and Its Application,* arXiv: physics/0504001 v1.
14. J.P. Bouchaud and R. Cont, Eur. Phys. J. B 6 543 (1998).
15. J. D. Farmer, adap-org/9812005.
16. J. Lakonishok, A. Shleifer, R. Thaler and R. Vishny, J. Fin. Econ. 32, 23 (1991).
17. Kephart, J. O., Hogg, T. & Huberman, B. A. *Dynamics of computational ecosystems.* Physical Review A 40(1), 404-421 (1989).